\documentclass[a4paper,11pt]{article}
\pdfoutput=1
\usepackage{jheppub}

\usepackage[T1]{fontenc}

\usepackage{url}
\usepackage{empheq}
\usepackage{amsmath}
\usepackage{amssymb}
\usepackage{dsfont}
\usepackage{graphicx}
\usepackage{hyperref}
\usepackage{stackrel}
\usepackage{cancel}
\usepackage[utf8]{inputenc}  
\usepackage{tikz}
\usetikzlibrary{patterns,arrows,positioning,shapes.misc,decorations.pathmorphing}

\bibstyle{JHEP}

\newcommand{\cO}{\mathcal{O}}
\newcommand{\DeltaH}{\Delta\corr{H_0}}

\DeclareMathOperator{\tr}{tr}

\newcommand{\bra}[1]    {\langle{#1}\vert}
\newcommand{\ket}[1]    {\vert{#1}\rangle}

\newcommand{\corr}[1]{\left\langle{#1}\right\rangle}

\newcommand{\be}    {\begin{equation}}
\newcommand{\ee}    {\end{equation}}

\newcommand{\Sec}{Section }
\newcommand{\Fig}{Figure }

\begin{document}

\title{Properties of Modular Hamiltonians on Entanglement Plateaux}

\author[a]{Raimond Abt}
\author{and}
\author[a]{Johanna Erdmenger}

\affiliation[a]{Institut f{\"u}r Theoretische Physik und Astrophysik,\\ Julius-Maximilians-Universit{\"a}t W{\"u}rzburg, Am Hubland, 97074 W\"urzburg, Germany}

\abstract{
The modular Hamiltonian of reduced states, given essentially by the
logarithm of the reduced density matrix,  plays an important role
within the AdS/CFT correspondence in view of its relation to quantum information. In particular, it is an essential
ingredient for quantum information measures of distances between
states, such as the relative entropy and the Fisher information metric.
However, the modular Hamiltonian
is known explicitly only for a few examples. For a family of states
$\rho_\lambda$ that is parametrized by a scalar $\lambda$, the first
order contribution
in $\tilde\lambda=\lambda-\lambda_0$ of the modular Hamiltonian to the
relative entropy between $\rho_\lambda$ and a reference state
$\rho_{\lambda_0}$ is completely determined by the entanglement
entropy, via the first law of entanglement. For several examples,
e.g.~for ball-shaped regions in the ground state of CFTs,  higher order
contributions are known to vanish. In these cases the modular Hamiltonian contributes to the Fisher information metric in a trivial way. We investigate under which conditions
the modular
Hamiltonian provides a non-trivial contribution to the Fisher information metric, i.e.~when the contribution of the modular Hamiltonian to the 
relative entropy is of higher order in $\tilde{\lambda}$.
We consider one-parameter families of reduced states on two entangling
regions that form an entanglement plateau, i.e.~the entanglement
entropies of the two regions saturate the Araki-Lieb inequality.
We show that in general, at least one of the relative entropies of the
two entangling regions is expected to involve
$\tilde{\lambda}$ contributions of higher order from the modular Hamiltonian.
Furthermore, we consider the implications of this observation for
prominent AdS/CFT examples that form entanglement plateaux in the large
$N$ limit. These examples include black strings with two sufficiently
close intervals, which we then generalize to an arbitrary number of
intervals. Moreover, we consider black branes with a  spherical shell,
as well as  BTZ black holes with large entangling intervals.
}


\maketitle

\section{Introduction}
\label{sec: Introduction}
One aspect of the AdS/CFT correspondence that caught significant
attention recently
 is its relation to quantum information (QI). The most prominent discovery in this field is the seminal
Ryu-Takayanagi (RT) formula \cite{2006PhRvL..96r1602R},
\begin{equation}
\label{eq: RT}
S(A)
=
\frac{\text{area}(\gamma_A)}{4G_N}\,.
\end{equation}
It relates the entanglement entropy $S$ of an
entangling region $A$ on the CFT side to the area of a minimal bulk surface $\gamma_A$ in the large $N$ limit. $\gamma_A$ is referred to as RT surface and
$G_N$ is Newton's constant. Starting from the RT formula, major progress was made in understanding the QI aspects of the field theory side
by studying the bulk. Further prominent examples for gravity dual
realizations of quantities relevant for QI are 
quantum error correcting codes \cite{Pastawski:2015qua}, the \textit{Fisher information metric} (FIM) \cite{Lashkari:2015hha, Banerjee:2017qti} and complexity \cite{Susskind:2014rva, Stanford:2014jda, Brown:2015bva}.
In particular, subregion complexity was proposed to be related
to the volume enclosed by RT surfaces \cite{Alishahiha:2015rta}. This
volume was recently related to a field-theory expression in \cite{Abt:2017pmf,Abt:2018ywl}.

In this paper we focus on the \textit{modular Hamiltonian} $H$ for general QFTs, which is defined by
\begin{equation}
	\rho
	=
	\frac{e^{-H}}{\tr(e^{-H})} \label{modham}
\end{equation}
for a given state $\rho$. \footnote{We use the terms density matrix
  and states interchangeably.} The modular Hamiltonian plays
an important role for QI measures such as the \textit{relative entropy}
(RE) (see e.g.~\cite{RevModPhys.74.197, Jafferis:2015del, Sarosi:2016oks, Sarosi:2016atx})  or the FIM and was studied comprehensively by many authors,
for instance in \cite{Wong:2013gua, Jafferis:2014lza, Lashkari:2015dia, Faulkner:2016mzt, Ugajin:2016opf, Arias:2016nip, Koeller:2017njr, Casini:2017roe, Sarosi:2017rsq, Arias:2017dda}. Many
interesting aspects of the modular Hamiltonian were investigated, such
as a quantum version of the Bekenstein bound \cite{Casini:2008cr,
  Blanco:2013lea} or a topological condition under which the modular
Hamiltonian of a 2d CFT can be written as a local integral over the
energy momentum tensor multiplied by a local weight
\cite{Cardy:2016fqc}. However,
the modular Hamiltonian is known explicitly only for a few examples,
such as for the reduced CFT ground
state on a ball-shaped entangling region in any dimension
(see e.g.~\cite{Casini:2011kv}) or for reduced thermal states on an interval for a $1+1$ dimensional CFT  (see e.g.~\cite{Lashkari:2014kda, Blanco:2017xef}).

This paper is devoted to determining further properties of the modular
Hamiltonian as given by \eqref{modham}, in particular in connection with an external variable $\lambda$ parametrizing the density matrix $\rho_\lambda$. This
parameter may be related to the energy density or the
temperature of the state, for instance, as we do in the examples considered below.
We obtain new results on the parameter dependence of
\begin{equation}
\label{eq: DeltaH def}
	\Delta\corr{H_0}(A,\lambda)
	=
	\tr(\rho_{\lambda}^A H_0)
	-
	\tr(\rho_{\lambda_0}^A H_0)\,,
\end{equation}
where $\rho_{\lambda}^A=\tr_{A^c}(\rho_{\lambda})$ is a reduced state on an entangling region $A$ and $H_0$ is the modular Hamiltonian of a chosen reduced reference state $\rho_{\lambda_0}^A$, i.e.
\begin{equation}
	\rho_{\lambda_0}^A
	=
	\frac{e^{-H_0}}{\tr(e^{-H_0})}\,.
\end{equation}
$\DeltaH$ plays a crucial role in the computation of the RE w.r.t.~$A$ of the one-parameter family of states $\rho_\lambda$,
\begin{equation}
\label{eq: Srel}
	S_{rel}(A,\lambda)
	=
	\tr(\rho_\lambda^A\log \rho_\lambda^A)
	-
	\tr(\rho_{\lambda}^A\log \rho_{\lambda_0}^A)
	=
	\Delta\corr{H_0}(A,\lambda)
	-
	\Delta S(A,\lambda)\,,
\end{equation}
as well as the FIM
\begin{equation}
\label{eq: Fisher info}
	G_{\lambda\lambda}(A,\lambda_0)
	=
	\partial_\lambda^2 S_{rel}(A,\lambda)|_{\lambda=\lambda_0}\,,
\end{equation}
where $\Delta S(A,\lambda)=S(A,\lambda)-S(A,\lambda_0)$ is the difference of the entanglement entropies $S(A,\lambda)$ and $S(A,\lambda_0)$ of the reduced states $\rho_\lambda^A$ and $\rho_{\lambda_0}^A$, respectively.
In particular for holographic theories, where the entanglement entropy is given by the RT formula \eqref{eq: RT}, $\DeltaH$ is the term that makes it difficult to
compute the RE and the FIM.
From \eqref{eq: Fisher info} we see however that $\DeltaH$ does not affect
$G_{\lambda\lambda}$ if it has at most linear contributions in $\lambda$. So
in these situations an explicit expression for $\DeltaH$ is not required to compute
$G_{\lambda\lambda}$.
We investigate the case when $\DeltaH$ contributes to the FIM in a non-trivial way, i.e.~when higher order $\lambda$ contributions are present in
$\DeltaH$. Since
\begin{equation}
	\DeltaH(A,\lambda_0)
	=
	0\,,
\end{equation}
from now on we refer to higher order contributions in $\tilde{\lambda}=\lambda-\lambda_0$
instead of $\lambda$.
\\

We examine the $\tilde{\lambda}$ dependence of $\DeltaH$ by considering the RE,
which is a valuable quantity for studying the modular Hamiltonian \cite{Casini:2008cr, Blanco:2013joa, Blanco:2013lea, Blanco:2017akw}.
For instance, the RE is known to be non-negative and to vanish iff $\rho^A_\lambda=\rho^A_{\lambda_0}$, which implies the \textit{first law of entanglement} \cite{Blanco:2013joa},
\begin{equation}
\label{eq: 1st law}
	\partial_\lambda\DeltaH(A,\lambda)|_{\lambda=\lambda_0}
	=
	\partial_\lambda\Delta S(A,\lambda)|_{\lambda=\lambda_0}\,.
\end{equation}
We see that even though the modular Hamiltonian $H_0$ is not known in general, we may use the the non-negativity of $S_{rel}$ to determine the leading order contribution of $\DeltaH$ in $\tilde{\lambda}$,
\begin{equation}
\label{eq: mod Ham taylor}
	\DeltaH(A,\lambda)
	=
	\partial_\lambda\Delta S(A,\lambda)|_{\lambda=\lambda_0}\tilde{\lambda}
	+
	\cO(\tilde{\lambda}^2)\,.
\end{equation}

For some configurations, such as thermal states dual to black string geometries with the energy density as parameter $\lambda$ and an arbitrary interval as entangling region $A$ \cite{Lashkari:2014kda, Blanco:2017xef}, the higher-order contributions in $\tilde{\lambda}$ are known to vanish\footnote{We discuss this setup in \Sec \ref{sec: Black strings}.}, i.e.
\begin{equation}
\label{eq: linear mod Ham}
	\DeltaH(A,\lambda)
	=
	\partial_\lambda\Delta S(A,\lambda)|_{\lambda=\lambda_0}\tilde{\lambda}\,.
\end{equation}
Consequently, $\DeltaH$ is completely determined by entanglement entropies, and in particular only contributes trivially to the FIM, as discussed above.
However, in general higher-order contributions in $\tilde{\lambda}$ will be present.
\\

In this paper we introduce a further application of the RE that allows us to
determine under which conditions higher-order contributions in $\tilde{\lambda}$ to $\DeltaH$ are to be
expected for families of states that form so-called \textit{entanglement plateaux}.
The term entanglement plateau was first introduced in \cite{Hubeny:2013gta} and
refers to entangling regions $A$, $B$ that saturate the \textit{Araki-Lieb inequality} (ALI) \cite{araki1970}
\begin{equation}
\label{eq: ALI}
	|S(A)-S(B)|\leq S(AB)\,.
\end{equation}

\begin{figure}[t]
\begin{center}
\includegraphics[scale=0.1]{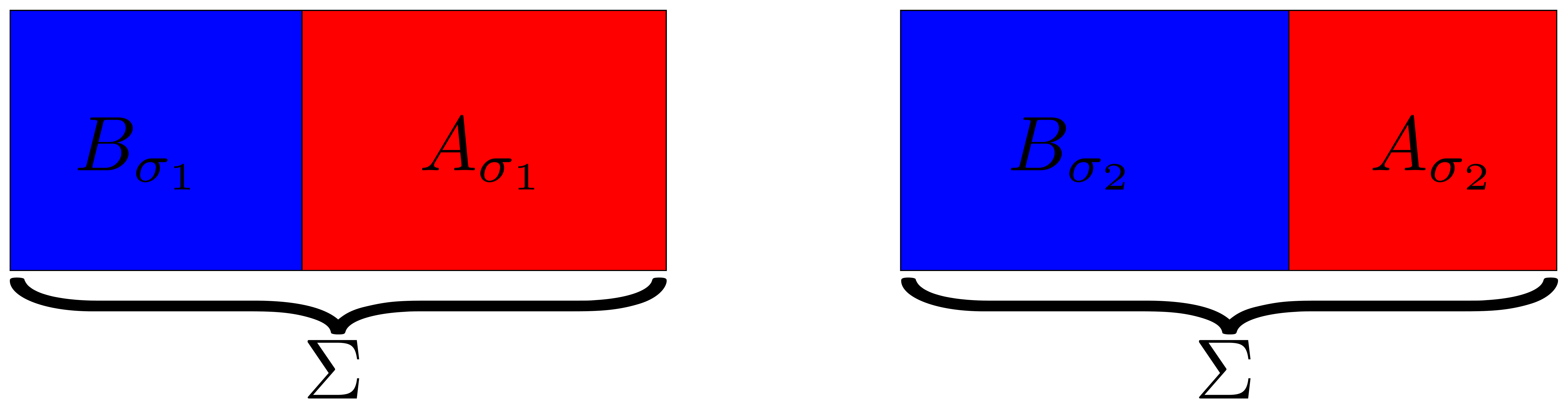} 
\end{center}
\caption{The families of entangling regions $A_\sigma$ and $B_\sigma$.
	We consider two families of entangling regions $A_\sigma$ (red) and $B_\sigma$ (blue)
	with $A_{\sigma_2}\subset A_{\sigma_1}$ for $\sigma_1< \sigma_2$ and
	$A_\sigma B_\sigma=\Sigma=const.$ In particular, this implies $B_{\sigma_1}\subset B_{\sigma_2}$.}
\label{fig: setup}
\end{figure}

We focus on entanglement plateaux that are stable under variations of $A$ and $B$ that keep $AB$ fixed. To be more precise, we consider two families $A_\sigma$ and $B_\sigma$
of entangling regions that come with a continuous parameter $\sigma$
determining their size, where $A_{\sigma_2}\subset A_{\sigma_1}$ if
$\sigma_1<\sigma_2$ and $A_{\sigma}B_{\sigma}=\Sigma=const.$ (see \Fig \ref{fig: setup}), and
saturate the ALI, i.e.
\begin{equation}
\label{eq: AL saturation}
	|S(A_\sigma,\lambda)-S(B_\sigma,\lambda)|= S(\Sigma,\lambda)\,.
\end{equation}
We show that the only way how both $\DeltaH(A_\sigma,\lambda)$ and $\DeltaH(B_\sigma,\lambda)$ can be linear in $\tilde{\lambda}$ for all $\sigma$ in a given interval $[\xi,\eta]$ is if $\partial_\lambda^2 S(A_\sigma,\lambda)$
and $\partial_\lambda^2 S(B_\sigma,\lambda)$ are constant in $\sigma$ on $[\xi,\eta]$.
The proof of this statement is a simple application of the well-known monotonicity \cite{uhlmann1977} of the RE,
\begin{equation}
\label{eq: monotonicity of Srel}
S_{rel}(A,\lambda)\leq S_{rel}(A',\lambda)\quad\text{if}\quad A\subseteq A'\,,
\end{equation}
and holds for any quantum system, not just for those with a holographic dual.
We thus find that in the setup described above, it suffices to look at
the entanglement entropies to see when higher-order contributions of $\tilde{\lambda}$
may be expected in at least one of the $\DeltaH$ (i.e.~$\DeltaH(A_\sigma,\lambda)$ or $\DeltaH(B_\sigma,\lambda)$), namely if $\partial_\lambda^2S(A_\sigma,\lambda)$ or $\partial_\lambda^2S(B_\sigma,\lambda)$ is not constant in $\sigma$.

In particular if one of the $\DeltaH$, say $\DeltaH(B_\sigma,\lambda)$,
is known to be linear for all $\sigma\in[\xi,\eta]$, we learn that $\DeltaH(A_\sigma,\lambda)$ is not.
Consequently, it is not sufficient to work with entanglement entropies to
determine $\DeltaH(A_\sigma,\lambda)$ via \eqref{eq: linear mod Ham}, but more involved calculations are required.
As a result this means that the RE is not just given by entanglement entropies.
\\

Our result for entanglement plateaux has important consequences in particular for
holographic theories. There are many well-known
configurations in holography that form entanglement plateaux in the large $N$ limit. Prominent examples -- which we discuss in this paper -- are large intervals for the BTZ back hole \cite{Headrick:2007km, Blanco:2013joa, Hubeny:2013gta} and two sufficiently close intervals for black strings \cite{Headrick:2010}. For these situations, very little is known about $\DeltaH$, \footnote{Note that the vacuum modular Hamiltonian of two intervals is known explicitly for the 2d CFTs of the massless free fermion \cite{Casini:2009vk, Arias:2018tmw} and the chiral free scalar \cite{Arias:2018tmw}. In this paper however, we consider thermal states in strongly coupled CFTs with gravity duals.} however our result can be used to prove that non-linear $\tilde{\lambda}$ contributions play a role in the $\DeltaH$ occurring in these models.
For the situation of two intervals described above, this may be used to show
that the modular Hamiltonian is not an integral over the energy momentum tensor
multiplied by a local scaling, as it is the case for one interval.
\\

This paper is structured as follows. In \Sec \ref{sec: Black strings}
we consider the special case of black strings as a motivation and to introduce the basic arguments required to verify our result, which we prove in \Sec \ref{sec: Theorem} in its full generality. We then present several situations where the result can be applied in \Sec \ref{sec: applications}. These include
an arbitrary number of intervals for thermal states dual to black strings,
a spherical shell for states dual to black branes, a sufficiently large entangling interval for states dual to BTZ black holes and primary excitations in a CFT with large central charge, defined on a circle. Furthermore, we discuss examples where the prerequisites of our result are not satisfied in \Sec \ref{sec: vacuum states for CFTs}.  Finally we make some concluding remarks in \Sec \ref{sec: conclusions}.

\section{A Simple Example: Black Strings}
\label{sec: Black strings}
Our result for modular Hamiltonians, as described in the introduction
and proved below in \Sec \ref{sec: Theorem}, may be applied to a vast
variety of situations. As an illustration, we begin by a simple example that introduces
the basic arguments for our result and demonstrates its
usefulness. This example involves thermal CFT states in $1+1$ dimensions of inverse temperature $\beta$ with black strings as gravity duals,
\begin{equation}
\label{eq: BS geometry}
	ds_{BS}^2
	=
	\frac{L^2}{z^2}\Big(
		-\frac{z_h^2-z^2}{z_h^2}dt^2
		+
		\frac{z_h^2}{z_h^2-z^2} dz^2
		+
		dx^2
						\Big)\, ,
\end{equation}
where $z=z_h$ is the location of the black string and $L$ is the AdS radius. The asymptotic boundary, where the CFT is defined, lies at $z=0$. The energy density
\begin{equation}
\label{eq: lambda ito beta}
	 \lambda
	 =
	 \frac{L}{16 \pi G_N z_h^2}
	 =
	 \frac{\pi c}{6 \beta^2}\,,
\end{equation}
where $c=\frac{3L}{2 G_N}$ is the central charge of the CFT,
is chosen as the parameter for this family of states. The reference
state may  be chosen to correspond to any energy density $\lambda_0$.
\\

We now demonstrate how the RE can be used to show that $\DeltaH$, as
defined in \eqref{eq: DeltaH def}, for a state living on two separated
intervals is in general not linear in $\tilde{\lambda}=\lambda-\lambda_0$ if the
two intervals are sufficiently close. The arguments that lead to this
conclusion will be  generalized in \Sec \ref{sec: Theorem} below.

\begin{figure}[t]
\begin{center}
\includegraphics[scale=0.2]{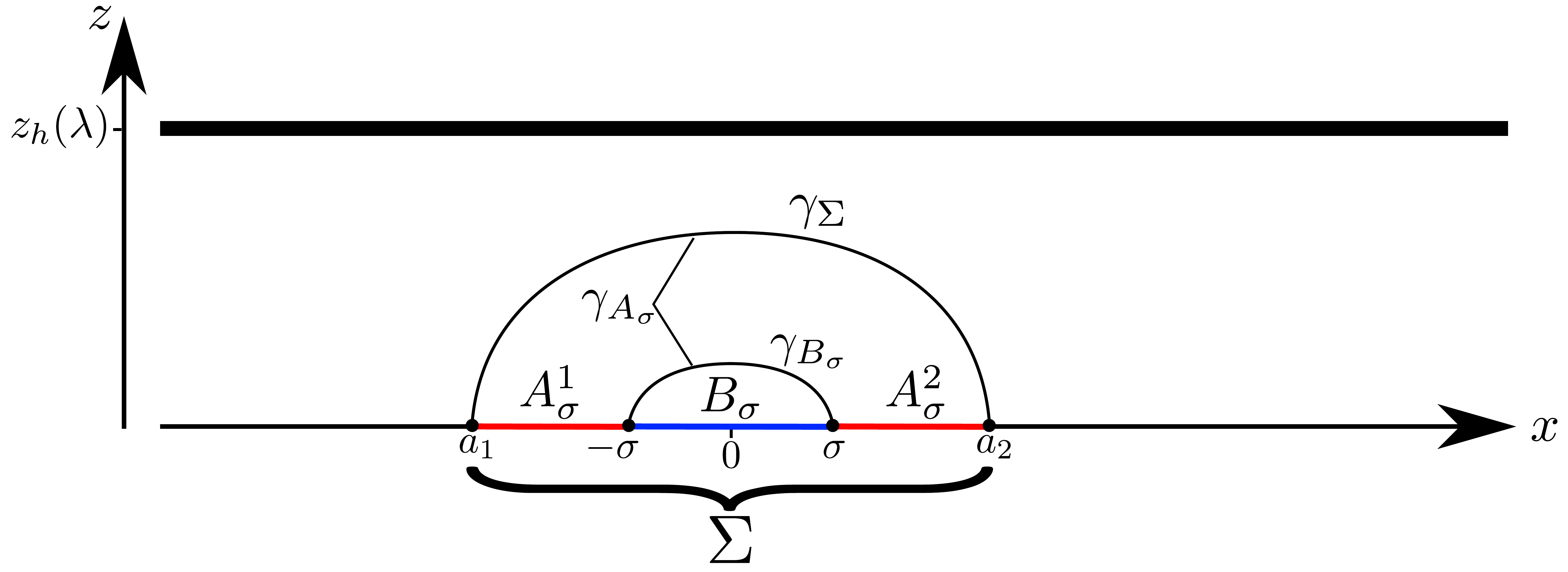} 
\end{center}
\caption{A constant time slice of the black string geometry \eqref{eq: BS geometry}. The asymptotic boundary of this geometry -- where the CFT is defined -- corresponds to the $x$-axis. The location of the black string is $z=z_h$ and depends on the energy density $\lambda$ via \eqref{eq: lambda ito beta}. If $\sigma$ is sufficiently small the RT surface $\gamma_{A_\sigma}$ of the entangling region $A_\sigma=A_\sigma^1A_\sigma^2$ (red) is the union of the RT surfaces $\gamma_\Sigma$ of $\Sigma=A_\sigma B_\sigma$ and $\gamma_{B_\sigma}$ of $B_\sigma$ (blue). This implies \eqref{eq: AL black string}.}
\label{fig: BS 2 int}
\end{figure}

Consider an entangling region $A_\sigma$ that consists of two intervals $A^1_\sigma=[a_1,-\sigma]$ and $A^2_\sigma=[\sigma, a_2]$, with $\sigma>0$ and $a_1$, $a_2$ fixed (see \Fig \ref{fig: BS 2 int}). The interval $B_\sigma=[-\sigma,\sigma]$ between $A^1_\sigma$ and $A^2_\sigma$ is w.l.o.g.~assumed to lie symmetric around the coordinate origin $x=0$.
If $\sigma$ is sufficiently small\footnote{For previous work regarding the modular Hamiltonian for such a situation, see e.g.~\cite{Blanco:2013joa}.},
the RT surface $\gamma_{A_\sigma}$ of $A_\sigma$ is the union of $\gamma_{B_\sigma}$ and $\gamma_\Sigma$ (see \Fig \ref{fig: BS 2 int}), where $\Sigma=A_\sigma B_\sigma=[a_1, a_2]$ is the union of $A_\sigma$ and $B_\sigma$.
Consequently, the entanglement entropy of $A_\sigma$ saturates the ALI \cite{Headrick:2010}, i.e.
\begin{equation}
\label{eq: AL black string}
	S(A_\sigma, \lambda)
	=
	S(\Sigma, \lambda)
	+
	S(B_\sigma, \lambda)\,,
\end{equation} 
which is an immediate consequence of the RT formula \eqref{eq: RT}.
For thermal states in general CFTs defined on the real axis, the modular Hamiltonian $H_0(B_\sigma)$ of $B_\sigma$ for the reference parameter value $\lambda_0$ is given by \cite{Lashkari:2014kda, Blanco:2017xef}
\begin{equation}
\label{eq: mod Ham one interval}
	H_0(B_\sigma)
	=
	\int_{-\sigma}^\sigma dx\, \beta_0\frac{\cosh(\frac{2\pi \sigma}{\beta_0})-\cosh(\frac{2\pi x}{\beta_0})}{\sinh(\frac{2\pi \sigma}{\beta_0})}T_{00}(x)\,,
\end{equation}
where $T_{\mu\nu}$ is the energy momentum tensor of the CFT and $\beta_0=\beta(\lambda_0)$. Thus, using \eqref{eq: DeltaH def}, we find
\begin{equation}
\label{eq: Delta mod Ham one interval}
	\DeltaH(B_\sigma,\lambda)
	=
	\beta_0\Big(2\sigma\coth\Big(\frac{2\pi \sigma}{\beta_0}\Big)-\frac{\beta_0}{\pi}\Big)\tilde{\lambda}
	=
	\Delta S'(B_\sigma,\lambda_0)\tilde{\lambda}
\end{equation}
to be linear in $\tilde{\lambda}$.
Here, the $'$ refers to a  derivative w.r.t.~$\lambda$.
The second equality is an immediate consequence of the first law of entanglement, i.e.~\eqref{eq: mod Ham taylor},
however may also be verified by a direct calculation using \cite{2006PhRvL..96r1602R, Calabrese:2004eu}
\begin{equation}
\label{eq: EE one interval}
	S(B_\sigma,\lambda)
	=
	\frac{c}{3}\log\Big(\frac{\beta}{\pi \epsilon}\sinh\Big(\frac{2\pi \sigma}{\beta}\Big)\Big)\ ,
\end{equation}
where $\epsilon$ is a UV cutoff.

\begin{figure}[t]
\begin{center}
\includegraphics[scale=0.1]{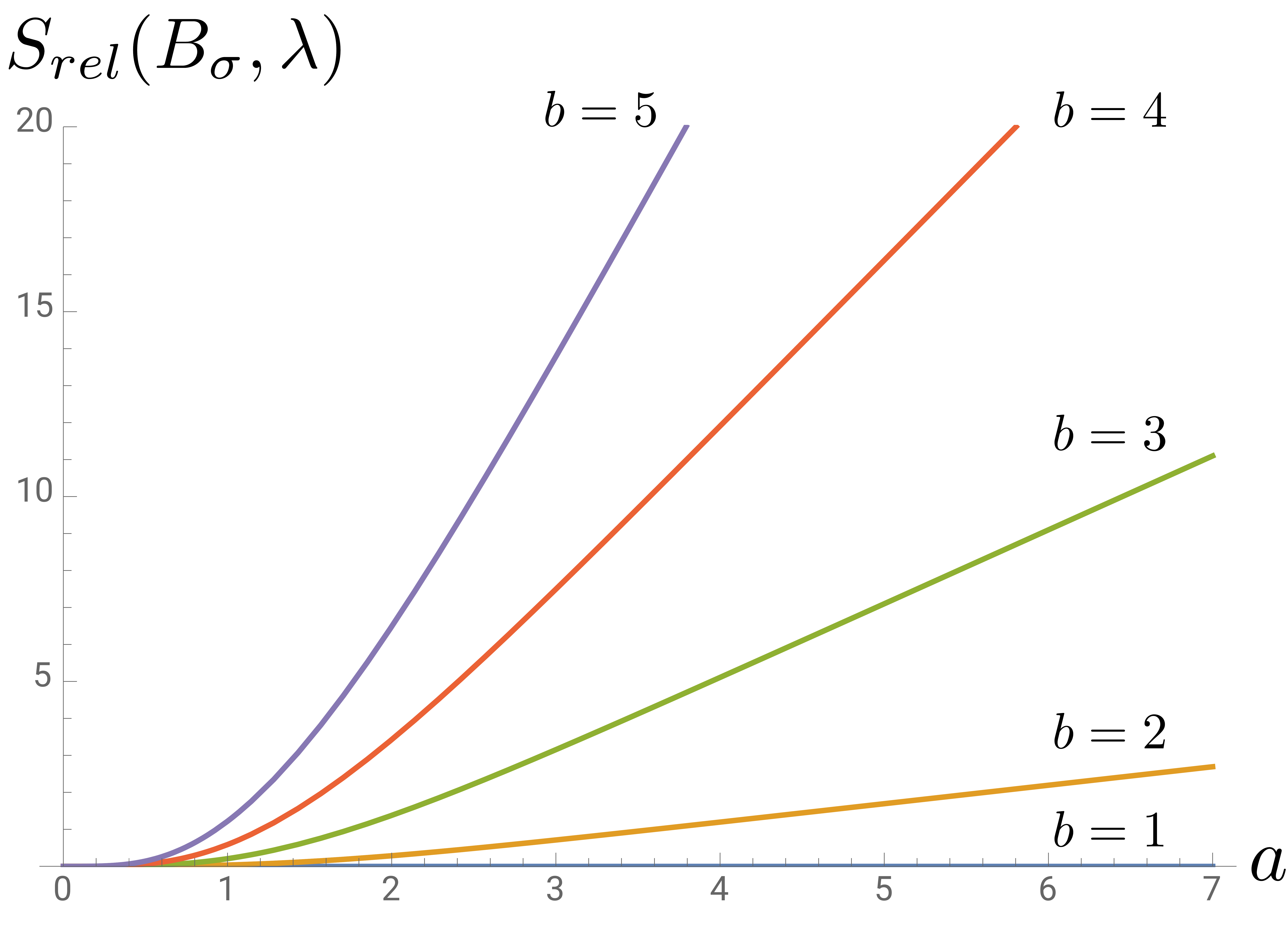} 
\end{center}
\caption{The behavior of $S_{rel}(B_\sigma, \lambda)$ \eqref{eq: Srel for B} w.r.t.~$a$ for $b=1,\dots,5$, where $a=2\pi\sigma/\beta_0$ and $b=\beta_0/\beta$. We set the global prefactor $c/3=1$ and see that $S_{rel}(B_\sigma, \lambda)$ grows with $a$ for fixed $b$. In particular, this implies that $S_{rel}(B_\sigma, \lambda)$ grows with $B_\sigma$, i.e.~$\sigma$, for fixed $\lambda$ and $\lambda_0$, which is in agreement with the monotonicity of the RE \eqref{eq: monotonicity of Srel}. For $b=1$ we find $S_{rel}(B_\sigma,\lambda)=0$, which is to be expected from \eqref{eq: Srel}, since this case corresponds to $\lambda=\lambda_0$.}
\label{fig: Srel plot}
\end{figure}

The two simple observations
\eqref{eq: AL black string} and \eqref{eq: Delta mod Ham one interval} together with the monotonicity of the RE \eqref{eq: monotonicity of Srel} are sufficient to verify that
$\DeltaH(A_\sigma,\lambda)$ is not linear in $\tilde{\lambda}$, except for possibly one particular $\sigma$, as we now show.
Let us assume that $\DeltaH(A_\sigma,\lambda)$ is linear in $\tilde{\lambda}$
for a given $\sigma$. The first law of entanglement \eqref{eq: mod Ham taylor} implies
\begin{equation}
	\DeltaH(A_\sigma,\lambda)
	=
	\Delta S'(A_\sigma,\lambda_0)\tilde{\lambda}\ .
\end{equation}
Applying this result to $S_{rel}(A_\sigma,\lambda)$ and using \eqref{eq: AL black string} and \eqref{eq: Delta mod Ham one interval},
we obtain
\begin{equation}
\label{eq: Srel two intervals}
	S_{rel}(A_\sigma,\lambda)
	=
	\Delta S'(\Sigma,\lambda_0)\tilde{\lambda}
	-
	\Delta S(\Sigma,\lambda)
	+
	S_{rel}(B_\sigma,\lambda)\ .
\end{equation}
Using \eqref{eq: lambda ito beta}, \eqref{eq: Delta mod Ham one interval} and \eqref{eq: EE one interval}, $S_{rel}(B_\sigma,\lambda)$ may be brought into the form
\begin{equation}
\label{eq: Srel for B}
	S_{rel}(B_\sigma,\lambda)
	=
	\frac{c}{3}\Big(
		\frac{1}{2}(1-b^2)(1-a\coth(a))
		+
		\log\Big(b\frac{\sinh(a)}{\sinh(b\,a)}\Big)
		\Big)\,,
\end{equation}
where $a=2\pi\sigma/\beta_0$ and $b=\beta_0/\beta$.
For fixed $b$, $S_{rel}(B_\sigma,\lambda)$ grows with $a$ (see \Fig \ref{fig: Srel plot}), which implies
that $S_{rel}(B_\sigma,\lambda)$ grows with $\sigma$ for fixed $\beta$ and $\beta_0$, or equivalently for fixed $\lambda$ and $\lambda_0$ (see \eqref{eq: lambda ito beta}).
Since $S_{rel}(B_\sigma,\lambda)$ is the only $\sigma$-dependent term on the RHS of \eqref{eq: Srel two intervals}, $S_{rel}(A_\sigma,\lambda)$ grows with $\sigma$ as well.
Now assume there were two values $\xi$, $\eta$ for $\sigma$, where we set w.lo.g. $\xi<\eta$, for which $\DeltaH(A_\sigma,\lambda)$ is linear in $\tilde{\lambda}$.
From the above discussion we conclude
\begin{equation}
	S_{rel}(A_\xi,\lambda)<S_{rel}(A_\eta,\lambda)\,.
\end{equation}
However, the monotonicity of the RE \eqref{eq: monotonicity of Srel} implies that $S_{rel}(A_\eta,\lambda)$ must be smaller than $S_{rel}(A_\xi,\lambda)$, since $A_\eta\subset A_\xi$.
So by assuming $\DeltaH(A_\sigma,\lambda)$ to be linear in $\tilde{\lambda}$ for more than one value of $\sigma$, we
encounter a contradiction. Consequently, $\DeltaH(A_\sigma,\lambda)$ may be linear in $\tilde{\lambda}$ for at most one particular $\sigma$.
\\

This simple example shows that  even though the modular
Hamiltonian for two disconnected intervals is unknown, general
properties of the RE imply that the modular Hamiltonian necessarily involves contributions of higher order in $\tilde{\lambda}$. An immediate consequence of this
observation is that the modular Hamiltonian for two intervals, unlike for one interval \eqref{eq: mod Ham one interval}, can \textit{not} be of the simple form
\begin{equation}
	\int_{A_\sigma} dx f^{\mu\nu}(x)T_{\mu\nu}(x)\,,
\end{equation}
where $f^{\mu\nu}$ is a local weight function, since this would lead to a $\DeltaH(A_\sigma,\lambda)$ that is linear in $\tilde{\lambda}$.
\\

Note that since $S_{rel}(B_\sigma,\lambda)$ is known, we are not required to consider
$\partial^2_\lambda S$, i.e.~the quantity discussed below \eqref{eq: monotonicity of Srel} in the introduction. We were able to deduce the non-linearity of $\DeltaH(A_\sigma,\lambda)$ directly from $S_{rel}(B_\sigma,\lambda)$ (see \eqref{eq: Srel two intervals}). In the more general cases discussed in \Sec \ref{sec: Theorem}, where $S_{rel}(B_\sigma,\lambda)$ is not known, this is no longer possible.
\section{Generic Entanglement Plateaux}
\label{sec: Theorem}
We now generalize the approach introduced in
\Sec \ref{sec: Black strings} and show how the RE  determines
whether non-linear contributions to $\DeltaH$ in $\tilde{\lambda}$ are to be
expected. Note that we do not require $\lambda$ to be the energy density, it is
just the variable that parametrizes the family of states
$\rho_\lambda$ we consider.

The discussion in \Sec \ref{sec: Black strings} required the saturation of the ALI \eqref{eq: ALI}, which allowed
us to show that if $\DeltaH$ were linear in $\tilde{\lambda}$, the RE would increase when the size of the considered entangling region (i.e.~two intervals) decreases. However, due to the monotonicity of the RE \eqref{eq: monotonicity of Srel} this is not possible.

By looking at \eqref{eq: Srel two intervals}, we see that this contradiction does not require the explicit expressions for the (relative) entropies: If $S_{rel}(B_\sigma,\lambda)$ grows with $B_\sigma$ for fixed $\Sigma$, $S_{rel}(A_\sigma,\lambda)$ grows as well. However, this is not compatible with the monotonicity of $S_{rel}$, since $A_\sigma=\Sigma\backslash B_\sigma$ decreases if $B_\sigma$ increases. This fact allows us to generalize the arguments of \Sec \ref{sec: Black strings} to generic entanglement plateaux, i.e.~systems that saturate the ALI.

\subsection{Result for Generic Entanglement Plateaux}
\label{sec: gen EP}
In the general case, the prerequisites for our main statement are as follows. We consider a one-parameter family of states $\rho_\lambda$. Let $\Sigma$ be an entangling region and $A_\sigma\subseteq\Sigma$ a one-parameter family of decreasing subregions of $\Sigma$, i.e.~$A_{\sigma_2}\subset A_{\sigma_1}$ for $\sigma_1<\sigma_2$, where the parameter $\sigma$ is assumed to be continuous. Furthermore, let $B_\sigma=\Sigma\backslash A_\sigma$ be the complement of $A_\sigma$ w.r.t.~$\Sigma$ (see \Fig \ref{fig: setup}).
Moreover, the ALI \eqref{eq: ALI} is assumed to be saturated for $A_\sigma$ and $B_\sigma$, i.e.
\begin{equation}
\label{eq: SA=SS+SB}
	|S(A_\sigma,\lambda)-S(B_\sigma,\lambda)|
	=
	S(\Sigma,\lambda)
	\quad
	\forall\sigma,\lambda\,.
\end{equation}
Furthermore, $S(A_\sigma,\lambda)$, $S(B_\sigma,\lambda)$ and $S(\Sigma,\lambda)$ are considered to be differentiable in $\lambda$ for all $\sigma$.

Subject to these prerequisites, we now state our main result.
If both $\DeltaH(A_{\sigma},\lambda)$ and $\DeltaH(B_{\sigma},\lambda)$ are linear in $\tilde{\lambda}=\lambda-\lambda_0$ for all $\sigma$ in a given interval $[\xi,\eta]$, then 
$\partial_\lambda^2S(A_\sigma,\lambda)$ and $\partial_\lambda^2S(B_\sigma,\lambda)$ are constant in $\sigma$ on $[\xi,\eta]$ for all $\lambda$.
\vspace{5mm}
\\
We prove this statement as follows.
As we discuss in the appendix, w.l.o.g.~we may restrict our arguments to the case $S(A_\sigma,\lambda)\geq S(B_\sigma,\lambda)$.
Assume that for all $\sigma\in[\xi,\eta]$, both $\DeltaH(A_{\sigma},\lambda)$ and $\DeltaH(B_{\sigma},\lambda)$ are linear in $\tilde{\lambda}$. Then, as explained in the introduction (see \eqref{eq: linear mod Ham}), we find 
\begin{equation}
\label{eq: linear modhams for A and B}
	\DeltaH(A_{\sigma},\lambda)
	=
	\Delta S'(A_{\sigma},\lambda_0)\tilde{\lambda}
	\quad\mbox{and}\quad
	\DeltaH(B_{\sigma},\lambda)
	=
	\Delta S'(B_{\sigma},\lambda_0)
	\tilde{\lambda}\,,
\end{equation}
where $'$ again refers to a derivative w.r.t.~$\lambda$.
This implies together with \eqref{eq: Srel} and \eqref{eq: SA=SS+SB}
\begin{equation}
\label{eq: SrelA i.t.o. SrelB for alpha}
	S_{rel}(A_\sigma,\lambda)
	=
	\Delta S'(\Sigma,\lambda_0)\tilde{\lambda}
	-
	\Delta S(\Sigma,\lambda)
	+
	S_{rel}(B_\sigma,\lambda)\,.
\end{equation}
Due to the monotonicity \eqref{eq: monotonicity of Srel} of $S_{rel}$ we find
\begin{equation}
	S_{rel}(B_\xi,\lambda)\leq S_{rel}(B_\eta,\lambda)\,,
\end{equation}
since $B_\xi\subset B_\eta$. Using \eqref{eq: SrelA i.t.o. SrelB for alpha}, this implies
\begin{equation}
\label{eq: Srel inequality}
	S_{rel}(A_\xi,\lambda)\leq S_{rel}(A_\eta,\lambda)\,.
\end{equation}
By construction we have $A_\eta\subset A_\xi$. So the only way how  \eqref{eq: Srel inequality} may be compatible with the monotonicity of $S_{rel}$ is if $S_{rel}(A_\sigma,\lambda)$ is constant in $\sigma$ for $\sigma\in[\xi,\eta]$.
Thus by using \eqref{eq: Srel} and \eqref{eq: linear modhams for A and B}, we find
\begin{equation}
\label{eq: DDS=const}
	-\partial^2_\lambda S_{rel}(A_\sigma,\lambda)
	=
	-\partial^2_\lambda(
		\Delta S'(A_\sigma,\lambda_0)(\lambda-\lambda_0)
		-
		\Delta S(A_\sigma,\lambda)
		)
	=
	\partial^2_\lambda S(A_\sigma,\lambda)
\end{equation}
to be constant in $\sigma$ on $[\xi,\eta]$.

Due to \eqref{eq: SrelA i.t.o. SrelB for alpha} the fact that $S_{rel}(A_\sigma,\lambda)$ is constant in $\sigma$ for $\sigma\in[\xi,\eta]$ implies that $S_{rel}(B_\sigma,\lambda)$ is as well.
In an analogous way as for $A_\sigma$, we find $\partial_\lambda^2 S(B_\sigma,\lambda)$ to be constant in $\sigma$ on $[\xi,\eta]$.
This completes the proof of the general result stated at the beginning of this section.

\subsection{Discussion for Generic Entanglement Plateaux}
\label{sec: ge EP discussion}
In \Sec \ref{sec: gen EP} we presented our result for a generic situation where the ALI is saturated. Some comments are in order.

First we note that even though we presented an example from holography in \Sec \ref{sec: Black strings} as a motivation, we did not require holography at any point during the proof. Therefore our result is true for any quantum system.
\\

Furthermore, we required $\sigma$, i.e.~the parameter of the family of entangling regions $A_\sigma$, to be continuous, as can be read off the discussion in the appendix. However, if we in addition assume the sign of $S(A_\sigma,\lambda)-S(B_\sigma,\lambda)$ to be constant in $\sigma$, we can apply the result to discrete systems, such as spin-chains, as well. The proof works analogously as in the continuous case discussed in \Sec \ref{sec: gen EP}.
\\

In \Sec \ref{sec: gen EP} we showed that $\partial_\lambda^2 S(A_\sigma,\lambda)$ and $\partial_\lambda^2 S(B_\sigma,\lambda)$ beeing constant in $\sigma$ on an interval $[\xi,\eta]$ is a necessary condition for both $\DeltaH(A_\sigma,\lambda)$ and $\DeltaH(B_\sigma,\lambda)$ to be linear in $\tilde{\lambda}$ for all $\sigma\in[\xi,\eta]$. However, this condition is not sufficient, as we now demonstrate by presenting an example where $\partial_\lambda^2 S(A_\sigma,\lambda)$ and $\partial_\lambda^2 S(B_\sigma,\lambda)$ are constant in $\sigma$ but both $\DeltaH(A_\sigma,\lambda)$ and $\DeltaH(B_\sigma,\lambda)$ are not linear in $\tilde{\lambda}$.

We consider a free massless boson CFT in two dimensions defined on a circle with radius $\ell_{CFT}$. The family of states is
chosen to consist of exited states of the form
\begin{equation}
	\ket{\lambda}
	=
	e^{i\sqrt{2\lambda}\Phi}\ket{0}\,,
\end{equation}
where $\Phi$ is the boson field and $\ket{0}$ is the vacuum state.
We use their conformal dimension $(\lambda,0)$ to parametrize these states.
For the sake of this paper we assume the conformal dimension $\lambda$ to be
a continuous parameter\footnote{Note that the parameter $\lambda$ is assumed to be continuous in \Sec \ref{sec: gen EP}, since we take derivatives w.r.t.~it, e.g.~in \eqref{eq: SrelA i.t.o. SrelB for alpha}.}.
We define $A_\sigma$ to be an interval of angular size $2(\pi-\sigma)$ and
$B_\sigma=A_\sigma^c$ to be the complementary interval of angular size $2\sigma$.
Consequently, $\Sigma=A_\sigma B_\sigma$ is the entire circle and the fact that
$\ket{\lambda}$ is pure implies $S(\Sigma,\lambda)=0$ and $S(A_\sigma,\lambda)=S(B_\sigma,\lambda)$,
and therefore the saturation of the ALI \eqref{eq: ALI}.
The reference state $\ket{\lambda_0}$ can be chosen arbitrarily.
This setup was discussed in \cite{Lashkari:2015dia}, where the RE was found to be
\begin{align}
	\label{eq: RE bosons A}
	S_{rel}(A_\sigma,\lambda)
	&
	=
	(1+(\pi-\sigma)\cot(\sigma))\Big(\sqrt{2\lambda}-\sqrt{2\lambda_0}\Big)^2\,,
	\\
	\label{eq: RE bosons B}
	S_{rel}(B_\sigma,\lambda)
	&
	=
	(1-\sigma\cot(\sigma))\Big(\sqrt{2\lambda}-\sqrt{2\lambda_0}\Big)^2\,.
\end{align}
The author of \cite{Lashkari:2015dia} states that the entanglement entropies of $A_\sigma$
and $B_\sigma$ are constant in $\lambda$. Therefore, by applying \eqref{eq: Srel}
to \eqref{eq: RE bosons A} and \eqref{eq: RE bosons B} we find
\begin{align}
	\label{eq: H bosons A}
	\DeltaH(A_\sigma,\lambda)
	&
	=
	(1+(\pi-\sigma)\cot(\sigma))\Big(\sqrt{2\lambda}-\sqrt{2\lambda_0}\Big)^2\,,
	\\
	\label{eq: H bosons B}
	\DeltaH(B_\sigma,\lambda)
	&
	=
	(1-\sigma\cot(\sigma))\Big(\sqrt{2\lambda}-\sqrt{2\lambda_0}\Big)^2\,.
\end{align}
Obviously, both $\DeltaH(A_\sigma,\lambda)$ and $\DeltaH(B_\sigma,\lambda)$
are not linear in $\tilde{\lambda}=\lambda-\lambda_0$. However, since the entanglement entropy is constant in $\lambda$, we find $\partial_\lambda^2 S=0$
for $A_\sigma$ and $B_\sigma$, and therefore that $\partial_\lambda^2 S$ is constant in $\sigma$ for $A_\sigma$ and $B_\sigma$. Thus we see that
$\partial_\lambda^2 S$ being constant in $\sigma$ does not imply that both $\DeltaH(A_\sigma,\lambda)$ and $\DeltaH(B_\sigma,\lambda)$ are linear in $\tilde{\lambda}$.
Therefore it is a necessary but not a sufficient condition.
\\

The proof of our result presented in \Sec \ref{sec: gen EP} strongly relies on the first law of entanglement \eqref{eq: 1st law}. We need to emphasize that the
first law of entanglement only applies if the reference state corresponds to a parameter value $\lambda_0$ that is not a boundary point of the set of allowed parameter values $\lambda$. The fact that the first law of entanglement holds is a consequence of the non-negativity of $S_{rel}(A,\lambda)$ and $S_{rel}(A,\lambda_0)=0$. These two
properties imply that $S_{rel}$ is minimal at $\lambda=\lambda_0$ and therefore
we find
\begin{equation}
\label{eq: different 1sr law}
	\partial_\lambda S_{rel}(A,\lambda)|_{\lambda=\lambda_0}
	=
	0\,.
\end{equation}
Using \eqref{eq: Srel} it is easy to see that \eqref{eq: different 1sr law} is equivalent to the first law of entanglement. However, if $\lambda_0$ is a boundary point of the set of allowed $\lambda$, i.e.~if it is not possible to choose $\lambda<\lambda_0$ for instance, the minimality of $S_{rel}(A,\lambda_0)$ does not necessarily imply $\partial_\lambda S_{rel}(A,\lambda)|_{\lambda=\lambda_0}$ to vanish.

The free massless boson CFT we discuss above is an example for such a situation. Here the parameter $\lambda$ is the conformal dimension of the considered states and is therefore non-negative. By choosing the reference state to be the vacuum, i.e.~$\lambda_0=0$, \eqref{eq: RE bosons A} gives
\begin{equation}
	S_{rel}(A_\sigma,\lambda)
	=
	2(1+(\pi-\sigma)\cot(\sigma))\tilde{\lambda}\,,
\end{equation}
and therefore $\partial_\lambda S_{rel}(A_\sigma,\lambda)|_{\lambda=\lambda_0}\neq 0$.
Consequently, the first law of entanglement does not hold for this example.
Even though it has the expected properties according to our prediction, i.e.~both $\DeltaH(A_\sigma,\lambda)$ and $\DeltaH(B_\sigma,\lambda)$ are
linear in $\tilde{\lambda}$ and
$\partial_\lambda^2S(A_\sigma,\lambda)$ and $\partial_\lambda^2 S(B_\sigma,\lambda)$ are constant in $\sigma$ (see \eqref{eq: H bosons A}, \eqref{eq: H bosons B} for $\lambda_0=0$), the prerequisites of our result are not satisfied if the first law of entanglement does not hold.
\\

We only considered one-parameter families of states in \Sec \ref{sec: gen EP}.
However, our result can be straightforwardly generalized to an $n$-parameter family of states $\rho_\Lambda$ with $\Lambda=(\lambda^1,...,\lambda^n)$. The
reference state corresponds to $\Lambda=\Lambda_0=(\lambda^1_0,...,\lambda^n_0)$.
In an analogous way as for the one-parameter case we can show that the
only way how both $\DeltaH(A_\sigma,\Lambda)$ and $\DeltaH(B_\sigma,\Lambda)$
can be linear in $\Lambda-\Lambda_0$, i.e.~of the form\footnote{Here we use once more the first law of entanglement \eqref{eq: 1st law}.}
\begin{equation}
\begin{split}
	&
	\DeltaH(A_\sigma,\Lambda)
	=
	\partial_i\Delta S(A_\sigma,\Lambda)|_{\Lambda=\Lambda_0}(\lambda^i-\lambda_0^i)\,
	\\
	&
	\DeltaH(B_\sigma,\Lambda)
	=
	\partial_i\Delta S(B_\sigma,\Lambda)|_{\Lambda=\Lambda_0}(\lambda^i-\lambda_0^i)\,,
\end{split}
\end{equation}
where $\partial_i=\partial/\partial\lambda^i$, for all $\sigma\in[\xi,\eta]$ is if $\partial_i\partial_j S(A_\sigma,\Lambda)$
and $\partial_i\partial_j S(B_\sigma,\Lambda)$ are constant in $\sigma$ on
$[\xi,\eta]$.
\\
\subsection{Alternative Formulation}
\label{sec: alternative formulation}

For the examples we discuss in \Sec \ref{sec: applications}, it is more convenient to use the following alternative formulation of our result:

Consider the assumptions necessary for the result to be satisfied (see \Sec \ref{sec: gen EP}). If $\partial_\lambda^2 S(A_\sigma,\lambda)$ or $\partial_\lambda^2 S(B_\sigma,\lambda)$ is not constant in $\sigma$ on any interval $[\xi,\eta]$, then there are only single values of $\sigma$ where both $\DeltaH(A_\sigma,\lambda)$ and $\DeltaH(B_\sigma,\lambda)$
are linear in $\tilde{\lambda}$, i.e.~there is no interval $[\xi,\eta]$ where
both $\DeltaH(A_\sigma,\lambda)$ and $\DeltaH(B_\sigma,\lambda)$ are linear in $\tilde{\lambda}$ for all $\sigma\in[\xi,\eta]$.

In the original formulation, the linearity of $\DeltaH(A_\sigma,\lambda)$ and $\DeltaH(B_\sigma,\lambda)$ in $\tilde{\lambda}$ implies that the second derivative of the entanglement entropies of $A_\sigma$ and $B_\sigma$ are constant in $\sigma$. In the alternative formulation however, non-constancy in $\sigma$ of the second derivative of one of the entanglement entropies implies that in general $\DeltaH$ is non-linear in $\tilde{\lambda}$ for $A_\sigma$, $B_\sigma$ or both.
In the examples of \Sec \ref{sec: applications}, there are non-constant second derivatives of the entanglement entropies, and therefore the alternative formulation is more appropriate. 
\\

In the alternative formulation, the number of values for $\sigma$ where both $\DeltaH(A_\sigma,\lambda)$ and $\DeltaH(B_\sigma,\lambda)$ are linear in $\tilde{\lambda}$ is undetermined.
However, in \Sec \ref{sec: Black strings}, where we considered $A_\sigma$ to be the union of two intervals, we were able to show a stronger statement. We found that there is at most one such value for $\sigma$ and moreover, that $\DeltaH(A_\sigma,\lambda)$ is linear in $\tilde{\lambda}$ only for that value of $\sigma$.
The arguments of \Sec \ref{sec: Black strings} that lead to this conclusion
can be generalized to the case of generic entanglement plateaux if
\begin{equation}
\label{eq: Drel}
	D_{rel}(B_\sigma,\lambda)
	=
	\Delta S'(B_\sigma,\lambda_0)\tilde{\lambda}
	-
	\Delta S(B_\sigma,\lambda)
\end{equation}  
grows strictly monotonically with $\sigma$. In particular, if $\DeltaH(B_\sigma,\lambda)$ is known to be linear in $\tilde{\lambda}$, $D_{rel}(B_\sigma,\lambda)$ is the RE of $B_\sigma$, \footnote{This is an immediate consequence of the first law of entanglement \eqref{eq: 1st law}.} which is the case for the setup discussed in \Sec \ref{sec: Black strings}, for instance.

 Just as in \Sec \ref{sec: gen EP}, we assume w.l.o.g.~$S(A_\sigma,\lambda)\geq S(B_\sigma,\lambda)$. Under the assumption that there are two values $\xi$, $\eta$ for $\sigma$ where $\DeltaH(A_\sigma,\lambda)$ is linear in $\tilde{\lambda}$, we
find, analogous to the derivation of \eqref{eq: SrelA i.t.o. SrelB for alpha},
\begin{equation}
	S_{rel}(A_{\xi,\eta},\lambda)
	=
	\Delta S'(\Sigma,\lambda_0)\tilde{\lambda}
	-
	\Delta S(\Sigma,\lambda)
	+
	D_{rel}(B_{\xi,\eta},\lambda)\,.
\end{equation}
Since $D_{rel}(B_\sigma,\lambda)$ is assumed to grow strictly monotonically with $\sigma$, this implies for $\xi<\eta$
\begin{equation}
	S_{rel}(A_\xi,\lambda)<S_{rel}(A_\eta,\lambda)\,,
\end{equation}
which is not possible due to the monotonicity of $S_{rel}$ \eqref{eq: monotonicity of Srel}, since $A_\eta\subset A_\xi$. Consequently, there can only be one value of $\sigma$
where $\DeltaH(A_\sigma,\lambda)$ is linear in $\tilde{\lambda}$.
\section{Applications}
\label{sec: applications}

We now apply the general result of \Sec \ref{sec: gen EP} to holographic states dual to black strings, black branes and BTZ black holes. Moreover, we apply the result to pure states, which we first discuss in full generality and then consider primary excitations of a CFT with large central charge as an example. In all these configurations entanglement plateaux can be constructed, i.e.~situations where the ALI is saturated \eqref{eq: SA=SS+SB}, which is the only requirement for our result.

\subsection{Black Strings Revisited}

First we consider once more, as in \Sec \ref{sec: Black strings}, the situation of two sufficiently close intervals
for CFTs dual to black strings \eqref{eq: BS geometry}.
The parameter $\lambda$ is chosen to be the energy density \eqref{eq: lambda ito beta}.
We can confirm the conclusion we made in \Sec \ref{sec: Black strings} by applying the result of \Sec \ref{sec: gen EP}:

Using \eqref{eq: EE one interval} is easy to see that
$\partial_\lambda^2 S(B_\sigma,\lambda)$ is not constant in $\sigma$ on any interval. So the result of \Sec \ref{sec: gen EP} tells us
that there is no interval $[\xi,\eta]$ where both $\DeltaH(A_\sigma,\lambda)$ and $\DeltaH(B_\sigma,\lambda)$ are linear in $\tilde{\lambda}$ for all $\sigma\in[\xi,\eta]$. We know that
$\DeltaH(B_\sigma,\lambda)$ is linear in $\tilde{\lambda}$ for all $\sigma$
(see \eqref{eq: Delta mod Ham one interval}), and therefore conclude that
$\DeltaH(A_\sigma,\lambda)$ is not, except possibly for single values of $\sigma$.

From the discussion in \Sec \ref{sec: alternative formulation} we are even able to conclude that there is only one such $\sigma$.
This is due to the fact that $D_{rel}(B_\sigma,\lambda)$ \eqref{eq: Drel}, which is equal to
$S_{rel}(B_\sigma,\lambda)$ here, grows strictly monotonically with $\sigma$, as pointed out in \Sec \ref{sec: Black strings}.
This special value of $\sigma$ corresponds to the degenerate situation
where $B_\sigma$ vanishes and $A_\sigma$ becomes a single interval, i.e.~$\sigma=0$.

\begin{figure}[t]
\begin{center}
\includegraphics[scale=0.2]{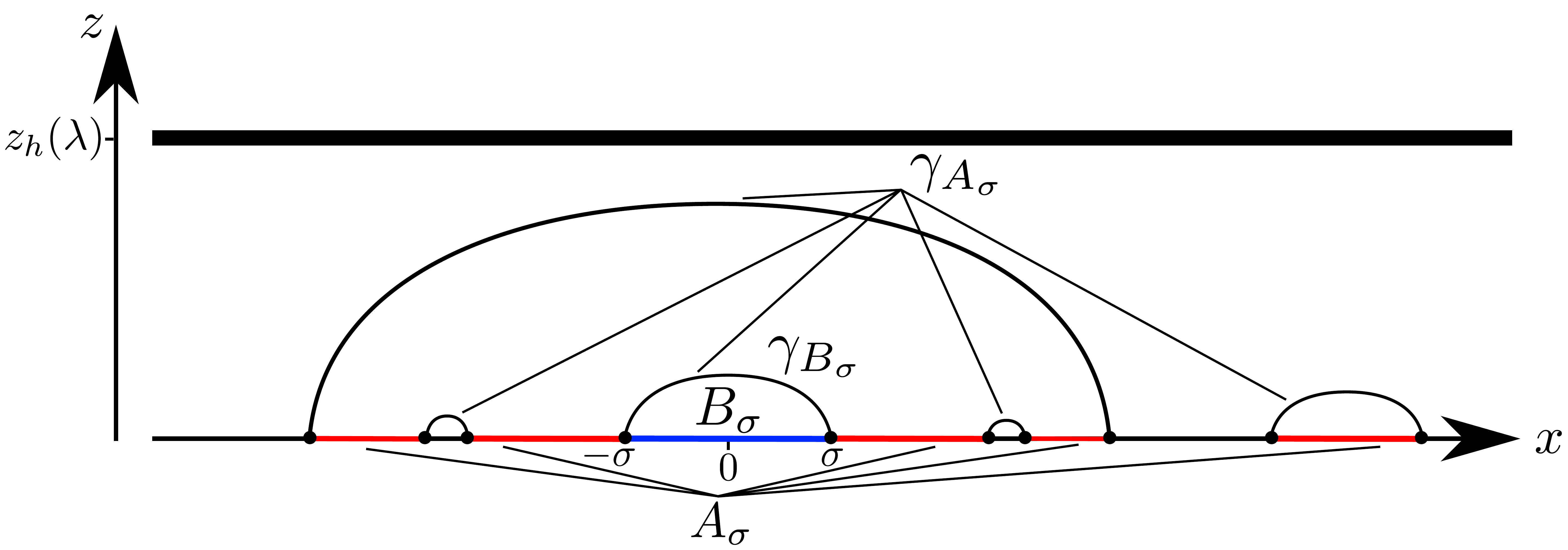} 
\end{center}
\caption{A constant time slice of the black string geometry \eqref{eq: BS geometry} revisited. The asymptotic boundary of this geometry -- where the CFT is defined -- corresponds to the $x$-axis. The location of the black string is $z=z_h$ and depends on the energy density $\lambda$ via \eqref{eq: lambda ito beta}. It is possible to choose a union of intervals $A_\sigma$ (red) and an interval $B_\sigma$ (blue) that lies between two intervals that belong to $A_\sigma$ in such a way that $A_\sigma$ and $B_\sigma$ saturate the ALI, i.e.~\eqref{eq: SA=SS+SB}.}
\label{fig: BS n int}
\end{figure}

The discussion of two intervals can be straightforwardly generalized to
the situation of $A_\sigma$ being the union of an arbitrary number of intervals. $B_\sigma$ is chosen to be an interval between two neighboring intervals that belong to $A_\sigma$. If the ALI is saturated, which corresponds to a situation such as the one depicted in \Fig \ref{fig: BS n int},
we see in analogy to the two-interval case, that $\DeltaH(A_\sigma,\lambda)$ is in general not linear in $\tilde{\lambda}$.

\subsection{Thermal States Dual to Black Branes}
\label{sec: black branes}
Consider thermal CFT states on $d$-dimensional Minkowski space that are dual to black branes,
\begin{equation}
	ds^2_{BB}
	=
	\frac{L^2}{z^2}\Big(
		-\frac{z_h^d-z^d}{z_h^d}dt^2
		+
		\frac{z_h^d}{z_h^d-z^d}dz^2
		+d\vec{x}_{d-1}^2\Big)\ ,
\end{equation}
where the black brane is located at $z=z_h$. Just as for black strings (see \Sec \ref{sec: Black strings}) the asymptotic boundary, where the CFT is defined, corresponds to $z=0$.
We choose $\Sigma$ to be a ball with radius $R$ and $B_\sigma$ another ball with radius $\sigma<R$  with the same center as $\Sigma$. Consequently, $A_\sigma=\Sigma\backslash B_\sigma$ is a spherical shell with inner radius $\sigma$ and outer radius $R$.
We choose $\lambda$ to be the energy density of the considered thermal states,
\begin{equation}
	\lambda
	=
	\frac{(d-1)L^{d-1}}{16 \pi G_N z_h^d}\,.
\end{equation}
The reference state is chosen to be the ground state, i.e.~$\lambda_0=0$. If we only consider sufficiently small radii $\sigma$, such that the RT surface of $A_\sigma$ is given by the union of the RT surfaces of $\Sigma$ and $B_\sigma$
for all $\sigma$, we find the ALI to be saturated for this setup (see \Fig \ref{fig: BS 2 int} for $d=2$).
Furthermore, we know $\DeltaH(B_\sigma,\lambda)$ to be linear in $\tilde{\lambda}$ for all $\sigma$ \cite{Blanco:2013joa},
\begin{equation}
\label{eq: H for BB}
	\DeltaH(B_\sigma,\lambda)
	=
	\frac{2\pi \Omega_{d-2}}{d^2-1}\sigma^d\tilde{\lambda}\,,
\end{equation}
where $\Omega_{d-2}=\frac{2\pi^{(d-1)/2}}{\Gamma((d-1)/2)}$.
Moreover, $S(B_\sigma,\lambda)$ is given, via the RT formula \eqref{eq: RT}, by \cite{Blanco:2013joa}
\begin{equation}
\label{eq: EE BB}
	S(B_\sigma,\lambda)
	=
	\frac{L^{d-1}\Omega_{d-2}}{4G_N}
	\int_0^\sigma d\rho \frac{\rho^{d-2}}{z(\rho)^{d-1}}\sqrt{1+\frac{(\partial_\rho z(\rho))^2 z_h^d}{z_h^d-z(\rho)^d}}
	\,,
\end{equation}
where $z(\rho)$ has to be chosen in such a way, that the integral on the RHS of
\eqref{eq: EE BB} is minimized. To our knowledge there is no analytic, integral free expression for $S(B_\sigma, \lambda)$ for generic $d$. However, in \cite{Blanco:2013joa} an expansion of $\Delta S(B_\sigma,\lambda)$ in
$\alpha\,\sigma^d\lambda$ is presented, with $\alpha=\frac{16\pi G_N}{d L^{d-1}}$, \footnote{As already pointed out in \cite{Sarosi:2016atx} there seems to be a typo in equation (3.55) of \cite{Blanco:2013joa}: The term $L^{d-1}/\ell_p^{d-1}$ needs to be inverted.}
\begin{equation}
\label{eq: Delta EE BB}
	\Delta S(B_\sigma,\lambda)
	=
	\frac{\Omega_{d-2}L^{d-1}}{4G_N}\Big(
		\frac{d\,\alpha\,\sigma^d\lambda}{2(d^2-1)}
		-
		\frac{d^3\sqrt{\pi}\,\Gamma(d-1)\alpha^2\sigma^{2d}\lambda^2}{
			2^{d+4}(d+1)\Gamma\Big(d+\frac{3}{2}\Big)}
		+
		\cO((\alpha\sigma^d\lambda)^3)\Big)\,.
\end{equation}
Due to $\partial_\lambda^2\Delta S(B_\sigma,\lambda)=\partial_\lambda^2S(B_\sigma,\lambda)$, we see that $\partial_\lambda^2S(B_\sigma,\lambda)$ is not constant in $\sigma$ on any interval. Since $\DeltaH(B_\sigma,\lambda)$ is linear in $\tilde{\lambda}$ \eqref{eq: H for BB} for all $\sigma$, the result of \Sec \ref{sec: gen EP} now tells us that $\DeltaH(A_\sigma,\lambda)$ may only be linear in $\tilde{\lambda}$ for single values of $\sigma$. \footnote{By applying our result to this situation we implicitly assume the first law of entanglement \eqref{eq: 1st law} to hold. However, as already pointed out in \cite{Blanco:2013joa} and \Sec \ref{sec: ge EP discussion}, the derivation of the first law for $\lambda_0=0$
would require to consider negative energy densities $\lambda<0$, which is unphysical. For the sake of this paper we assume the first law to be valid in the limit $\lambda_0\rightarrow 0$, since it holds for any $\lambda_0>0$.}

Just as for the black string, we can even show that there is only one such $\sigma$. From \eqref{eq: H for BB} and \eqref{eq: Delta EE BB} we conclude that $S_{rel}(B_\sigma,\lambda)$ \eqref{eq: Srel} is not constant in $\sigma$ on any interval. The monotonicity \eqref{eq: monotonicity of Srel} of the RE then implies that $S_{rel}(B_\sigma,\lambda)$ grows strictly monotonically with $\sigma$. Since $\DeltaH(B_\sigma,\lambda)$ is linear in $\tilde{\lambda}$ we find $D_{rel}(B_\sigma,\lambda)=S_{rel}(B_\sigma,\lambda)$ \eqref{eq: Drel} and therefore conclude that $D_{rel}(B_\sigma,\lambda)$ grows strictly monotonically with $\sigma$. The discussion in \Sec \ref{sec: alternative formulation} now implies that there is at most one value of $\sigma$ where $\DeltaH(A_\sigma,\lambda)$ is linear in $\tilde{\lambda}$. This special $\sigma$ can be found to be the degenerate case $\sigma=0$, i.e.~when $B_\sigma$ vanishes.

\subsection{BTZ Black Hole}
\label{sec: BTZ}

As a further application of the result of \Sec \ref{sec: gen EP} to holography we consider thermal states dual to BTZ black hole geometries,
\begin{equation}
\label{eq: BTZ geometry}
	ds_{BTZ}^2
	=
	-\frac{r^2-r_h^2}{L^2}dt^2
		+
	\frac{L^2}{r^2-r_h^2}dr^2
	+
	r^2 d\phi^2	
	\,.
\end{equation}
The horizon radius $r_h$ is given -- in terms of the CFT temperature $T$ and the radius $\ell_{CFT}$ of the circle on which the CFT is defined -- by
\begin{equation}
	r_h
	=
	\sqrt{8G_N M}L
	=
	2\pi L\ell_{CFT}T\,,
\end{equation}
where $M$ is the mass of the BTZ black hole.
\\

\begin{figure}[t]
\begin{center}
\includegraphics[scale=0.18]{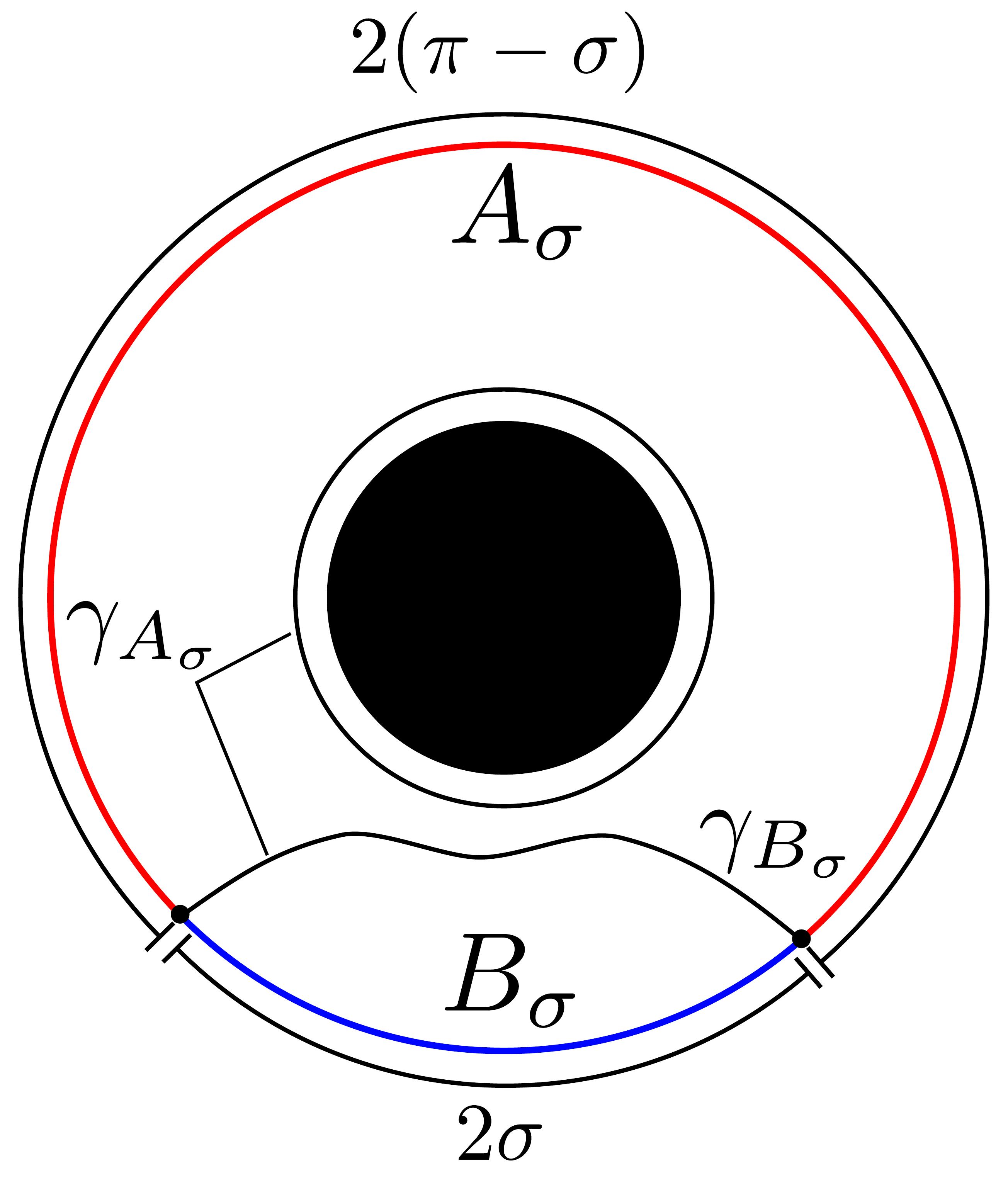} 
\end{center}
\caption{A constant time slice of the BTZ black hole geometry \eqref{eq: BTZ geometry}. The CFT is defined on the asymptotic boundary at $r\rightarrow \infty$. For a sufficiently large entangling region $A_\sigma$ (red) the RT surface $\gamma_{A_\sigma}$ is the union of the RT surface $\gamma_{B_\sigma}$ of its complement $B_\sigma$ (blue) and the black hole horizon. This implies \eqref{eq: SA=SS+SB}.}
\label{fig: BTZ}
\end{figure}

The asymptotic boundary, where the CFT is defined, corresponds to $r\rightarrow \infty$. For an interval $A_\sigma$ of sufficiently large angular size $2(\pi-\sigma)$, the RT surface consists of two disconnected parts: the horizon and the RT surface of $A^c_\sigma=B_\sigma$, as depicted in \Fig \ref{fig: BTZ}.
The entanglement entropy
is then given by \cite{Headrick:2007km, Blanco:2013joa}
\begin{equation}
\label{eq: EE BTZ}
	S(A_\sigma)
	=
	\frac{c}{3}2\pi^2T\ell_{CFT}
	+
	\frac{c}{3}\log\Big(\frac{1}{\pi T \epsilon}\sinh(
		2\pi\ell_{CFT}T\sigma)\Big)\,,
\end{equation}
where $\epsilon$ is a UV cutoff. The first term is the thermal entropy of the state and corresponds to the black hole horizon, while the second term is the
entanglement entropy of $B_\sigma$. We see once more that the states on $A_\sigma$ and $B_\sigma$ saturate the ALI.
As parameter $\lambda$ for this family of states we choose the square of the temperature,
\begin{equation}
\label{eq: lambda for BTZ}
	\lambda
	=
	T^2\,,
\end{equation}
which corresponds to the mass $M$ of the black hole,
\begin{equation}
\label{eq : T^2=M}
	LM
	=
	\frac{\pi^2\ell_{CFT}^2c}{3}\lambda\,.
\end{equation}
The reference state can be chosen to correspond to any $\lambda=\lambda_0=T_0^2$.
Using \eqref{eq: EE BTZ} it is straight forward to see that $\partial_\lambda^2 S(A_\sigma,\lambda)$ is not constant in $\sigma$ on any interval. So, even though the explicit forms of $\DeltaH(A_\sigma,\lambda)$ and $\DeltaH(B_\sigma,\lambda)$ \eqref{eq: DeltaH def} are not known, we can use the result of \Sec \ref{sec: gen EP} to conclude that in general at least one of $\DeltaH(A_\sigma,\lambda)$ or $\DeltaH(B_\sigma,\lambda)$ is not linear in $\tilde{\lambda}=T^2-T_0^2$.

Note that the result of \Sec \ref{sec: gen EP} cannot be used to determine whether $\DeltaH(A_\sigma, \lambda)$, $\DeltaH(B_\sigma, \lambda)$ or both are non-linear in $\tilde\lambda$. However, the discussion in \Sec \ref{sec: alternative formulation} actually allows us to show that $\DeltaH(A_\sigma,\lambda)$ is not linear in $\tilde\lambda$ for more than one particular $\sigma$: By applying
$S(B_\sigma,\lambda)$, i.e.~the second term in \eqref{eq: EE BTZ}, to \eqref{eq: Drel} we find
\begin{equation}
	D_{rel}(B_\sigma,\lambda)
	=
	\frac{c}{3}\Big(
		\frac{1}{2}\big(
			1-\tilde{a}\coth(\tilde{a})
			\big)(1-\tilde{b}^2)
		+
		\log\Big(\tilde{b}\frac{\sinh(\tilde{a})}{\sinh(\tilde{b}\,\tilde{a})}\Big)			
		\Big)\,,
\end{equation}
where $\tilde{a}=2\pi \ell_{CFT}\sqrt{\lambda_0}\sigma$ and $\tilde{b}=\sqrt{\lambda}/\sqrt{\lambda_0}$. The structure of the $\sigma$ dependence of $D_{rel}(B_\sigma,\lambda)$ is identical to the structure of the
$\sigma$ dependence of $S_{rel}(B_\sigma,\lambda)$ that was derived in \Sec \ref{sec: Black strings}
for two intervals (see \eqref{eq: Srel two intervals} and \eqref{eq: Srel for B}). 
So in an analogous way to the discussion in \Sec \ref{sec: Black strings}, we find that $D_{rel}(B_\sigma,\lambda)$ grows strictly monotonically with $\sigma$. Consequently,
$\DeltaH(A_\sigma,\lambda)$ is not linear in $\tilde{\lambda}$ except for possibly one particular $\sigma$.

\subsection{Pure States: Primary Excitations in CFTs with Large Central Charge}
\label{sec: pure states}

It is also possible to apply the result of \Sec \ref{sec: gen EP} to a one-parameter family of pure states. Consider $\rho_\lambda$ to be such a family and $\Sigma$ to be the entire constant time slice, i.e.~$B_\sigma=A_\sigma^c$. Since $S(\Sigma, \lambda)=0$ and $S(A_\sigma,\lambda)=S(B_\sigma,\lambda)$, the ALI is saturated for this setup. The result of \Sec \ref{sec: gen EP} now tells us that if $\partial_\lambda^2S(A_\sigma,\lambda)$ is not constant in $\sigma$ on any interval $[\xi,\eta]$, it is not possible for $\DeltaH(A_\sigma,\lambda)$
and $\DeltaH(B_\sigma,\lambda)$ to be linear in $\tilde{\lambda}$ for the same $\sigma$, except for single values of $\sigma$.
\\

As an example for such a family of pure states we consider spinless primary excitations $\ket{\lambda}$ in a CFT with large central charge $c$ defined on a circle with radius $\ell_{CFT}$.
We use the conformal dimension
\begin{equation}
\label{eq: lambda for primaries}
	(h_\lambda,\bar{h}_\lambda)
	=
	\Big(\frac{c\lambda}{24}, \frac{c\lambda}{24}\Big)
\end{equation}
to parametrize these states\footnote{We have introduced the multiplicative factor $c/24$ in the definition of $\lambda$ to simplify the formulae in this section.}
and assume $\ket{\lambda}$ to correspond to a heavy operator, i.e.~$\Delta_\lambda=h_\lambda+\bar{h}_\lambda=\cO(c)$. Moreover, we restrict our analysis to the case $\lambda<1$ and assume the spectrum of light operators, i.e.~operators with $\Delta=h+\bar{h}\ll c$, to be sparse. The entangling regions $\Sigma$ and $B_\sigma$ are chosen to be the entire circle and an interval with angular size $2\sigma<\pi$, respectively. Consequently, $A_\sigma=B_\sigma^c$ is an interval with angular size $2(\pi-\sigma)>\pi$. The reference state corresponds to an arbitrary value $\lambda_0$ of the parameter $\lambda$.

The entanglement entropy of $B_\sigma$ for this setup was computed in \cite{Asplund:2014coa},
\begin{equation}
\label{eq: EE primaries}
	S(B_\sigma,\lambda)
	=
	\frac{c}{3}\log\Big(\frac{2\ell_{CFT}}{\sqrt{1-\lambda}\,\epsilon}\sin\big(\sqrt{1-\lambda}\,\sigma\big)\Big)
	=
	S(A_\sigma,\lambda)\,,
\end{equation}
where $\epsilon$ is a UV cutoff. The second equality in \eqref{eq: EE primaries}
 is a consequence of the fact that $\ket{\lambda}$ is pure\footnote{Note that the expression for $S(B_\sigma,\lambda)$ in \eqref{eq: EE primaries} is not symmetric under the transformation $\sigma\mapsto \pi-\sigma$, as one would naively expect
from the purity of $\ket{\lambda}$. The reason for that is the fact that in the derivation of $S(B_{\sigma},\lambda)$ \cite{Asplund:2014coa} $2\sigma<\pi$ was applied.} and ensures that the ALI is saturated.
It is easy to see that $\partial_\lambda^2 S(B_\sigma,\lambda)$ is not constant
in $\sigma$ on any interval. Therefore the result of \Sec \ref{sec: gen EP} implies that there are only single values of $\sigma$ where both $\DeltaH(A_\sigma,\lambda)$ and $\DeltaH(B_\sigma,\lambda)$ are linear in $\tilde{\lambda}=\lambda-\lambda_0$.

Analogously to the discussion regarding BTZ black holes in \Sec \ref{sec: BTZ}, we can actually show that $\DeltaH(A_\sigma,\lambda)$ is in not linear in $\tilde{\lambda}$ for any $\sigma$ with possibly one exception.

\subsection{Vacuum States for CFTs on a Circle}
\label{sec: vacuum states for CFTs}
We would like to emphasize an interesting observation regarding a family of primary states $\ket{\lambda}$ for a CFT defined on a circle with radius $\ell_{CFT}$. We define the entangling intervals $A_\sigma$ and $B_\sigma$ and the parameter $\lambda$ as in \Sec \ref{sec: pure states}. However, we do not require the CFT
to have large central charge. Furthermore, we do not assume any restrictions regarding the spectrum.
The reference state is chosen to be the vacuum state, i.e.~$\lambda_0=0$.
Since $\ket{\lambda}$ is a family of pure states, the ALI is saturated, as pointed out in \Sec \ref{sec: pure states}.

In this section we show that
our result of \Sec \ref{sec: gen EP} may be used to arrange the considered families of states into three categories: families where $\partial_\lambda^2S(A_\sigma,\lambda)$ and $\partial_\lambda^2S(B_\sigma,\lambda)$ are constant in $\sigma$, families where the parameter $\lambda$ is not continuous, such that the reference value $\lambda_0=0$ is separated from the other parameter values,
and finally families where the first law of entanglement \eqref{eq: 1st law} does not hold. These categories are not mutually exclusive.

For the example considered in this section, it is possible to choose these three categories since both $\DeltaH(A_\sigma,\lambda)$ and $\DeltaH(B_\sigma,\lambda)$ are linear in $\tilde{\lambda}$ for all $\sigma$, as may  be seen as follows.
In general, the modular Hamiltonian $H_0(2\varsigma)$ for the
ground state of a CFT on a circle, restricted to an interval with angular size $2\varsigma$, is given by \cite{Blanco:2013joa}
\begin{equation}
\label{eq: mod Ham vac circle}
	H_0(2\sigma)
	=
	2\pi\ell_{CFT}^2\int_0^{2\varsigma} d\phi\frac{\cos(\phi-\varsigma)-\cos(\varsigma)}{\sin(\varsigma)}T_{00}\,.
\end{equation}
Using the CFT result
\begin{equation}
	\bra{\lambda}T_{00}\ket{\lambda}
	-
	\bra{0}T_{00}\ket{0}
	=
	\frac{c\tilde{\lambda}}{24\pi\ell_{CFT}^2}\,,
\end{equation}
we find from \eqref{eq: mod Ham vac circle} that
\begin{equation}
\label{eq: modHam primaries}
	\DeltaH(A_\sigma,\lambda)
	=
	\frac{c}{6}\Big(1+(\pi-\sigma)\cot(\sigma)\Big)\tilde{\lambda}
\end{equation}
and
\begin{equation}
	\DeltaH(B_\sigma,\lambda)
	=
	\frac{c}{6}\Big(1-\sigma\cot(\sigma)\Big)\tilde{\lambda}
\end{equation}
are linear in $\tilde{\lambda}$.

The first category of families corresponds to the case where all
prerequisites of our result of \Sec \ref{sec: gen EP} are satisfied. Both $\DeltaH(A_\sigma,\lambda)$ and $\DeltaH(B_\sigma,\lambda)$ are linear in $\tilde{\lambda}$ for all $\sigma$, so we conclude that $\partial_\lambda^2S$ is constant in $\sigma$ for both $A_\sigma$ and $B_\sigma$.

If $\partial_\lambda^2S(A_\sigma,\lambda)$
or $\partial_\lambda^2S(B_\sigma,\lambda)$ is not constant in $\sigma$, then at least one of the prerequisites of our result of \Sec \ref{sec: gen EP} is not satisfied. The examples with this property then fall into one of the other two categories introduced above.
 
There are two ways in which the prerequisites may be violated. One way is that the parameter $\lambda$ cannot be continuously continued to $\lambda_0=0$, which corresponds to the second category of families. In the proof of our result in \Sec \ref{sec: gen EP} we assume $\lambda$ to be continuous, since we take derivatives w.r.t.~$\lambda$ (see e.g.~\eqref{eq: SrelA i.t.o. SrelB for alpha}).
So if $\lambda$ has a gap at $\lambda_0$ the derivative w.r.t.~$\lambda$ is not
defined there.

The other way how the prerequisites may be violated is when the first law of entanglement does not hold. Since the conformal dimension is always non-negative, the reference value $\lambda_0=0$ is a boundary point of the set of allowed parameter values $\lambda$. As pointed out in \Sec \ref{sec: ge EP discussion}, the first law of entanglement may not apply in this case, since the first derivative of the RE may not vanish at $\lambda_0=0$. However, this law is an essential ingredient in the proof of \Sec \ref{sec: gen EP}. This situation corresponds to the third category of families.

To conclude, we note that our result of \Sec \ref{sec: gen EP} allows for a distinction of the three categories described in this section.
\section{Discussion}
\label{sec: conclusions}
In this paper we studied the modular Hamiltonian of a one-parameter family of reduced density matrices $\rho^{A,B}_\lambda$ on entangling regions $A$ and $B$ that form entanglement plateaux, i.e.~that saturate the ALI \eqref{eq: ALI}. These plateaux were considered to be stable under variations of $A$ and $B$ that leave $\Sigma=A B$ invariant. We parametrized these variations by introducing a continuous variable $\sigma$, i.e.~$A\rightarrow A_\sigma$, $B\rightarrow B_\sigma$, such that $A_{\sigma_2}\subset A_{\sigma_1}$ for $\sigma_1<\sigma_2$.

Our main result is that the only way how both $\DeltaH(A_\sigma,\lambda)$ and $\DeltaH(B_\sigma,\lambda)$, as defined in \eqref{eq: DeltaH def}, can be linear in $\tilde{\lambda}=\lambda-\lambda_0$ for all $\sigma$ in an interval $[\xi,\eta]$
is if $\partial_\lambda^2 S(A_\sigma, \lambda)$ and $\partial_\lambda^2 S(B_\sigma, \lambda)$ are constant in $\sigma$ on $[\xi,\eta]$. Subsequently to discussing this result for states dual to black strings as a motivation (see \Sec \ref{sec: Black strings}), we proved it in \Sec \ref{sec: gen EP} for arbitrary quantum systems using the first law of entanglement \eqref{eq: 1st law} and the monotonicity \eqref{eq: monotonicity of Srel} of the RE \eqref{eq: Srel}.

As we discussed in the introduction, if $\DeltaH$ is linear in $\tilde{\lambda}$
it effectively does not contribute to the FIM \eqref{eq: Fisher info}. So we see that in the
setup described above the FIM of $A_\sigma$, $B_\sigma$ or both will in general contain non-trivial contributions of $\DeltaH$.
Furthermore, if it is linear in $\tilde{\lambda}$, $\DeltaH$ is completely determined by the entanglement entropy via the first law of entanglement (see \eqref{eq: linear mod Ham}). In the setup described above however, we find that $\DeltaH(A_\sigma,\lambda)$, $\DeltaH(B_\sigma,\lambda)$ or both will in general
not have this simple form.

In \Sec \ref{sec: applications} we applied the result of \Sec \ref{sec: gen EP} to several prominent holographic examples of entanglement plateaux. By choosing $\lambda$ to be the energy density  of thermal states dual to black strings, we showed that higher-order contributions in $\tilde{\lambda}$ are present in $\DeltaH(A_\sigma,\lambda)$ for $A_\sigma$  being the union of two sufficiently close intervals. Furthermore, we showed a similar result for thermal states dual to black branes, where $\lambda$ was again chosen to be the energy density, $\lambda_0$ was set to $0$ and $A_\sigma$ was chosen to be a spherical shell with sufficiently small inner radius $\sigma$.
In these two situations, $\DeltaH(B_\sigma,\lambda)$ is known to be linear in $\tilde{\lambda}$. This allowed us to determine that $\DeltaH(A_\sigma,\lambda)$ must be non-linear in $\tilde{\lambda}$.

Moreover, we also discussed the BTZ black hole, where we chose $A_\sigma$ to be a sufficiently large entangling interval so that $A_\sigma$ and $B_\sigma=A_\sigma^c$ saturate the ALI. For this case we were able to use our result to show that at least one $\DeltaH(A_\sigma,\lambda)$ or $\DeltaH(B_\sigma,\lambda)$ is in general non-linear in $\tilde{\lambda}=T^2-T_0^2$, where $T$ is the CFT temperature. A more detailed analysis of the entanglement entropy even allowed us to determine that $\DeltaH(A_\sigma,\lambda)$ will have higher order $\tilde{\lambda}$ contributions.
We showed a similar result for primary excitations in a CFT on a circle with large central charge $c$. In this case $B_\sigma$ was set to be an interval with angular size $2\sigma<\pi$ and $A_\sigma=B_\sigma^c$. The parameter $\lambda$ was chosen to
be the conformal dimension multiplied by $c/24\pi$.
\\

We emphasize that even though all these examples are very different from each
other, the fact that non-linear contributions in $\tilde{\lambda}$ are to be expected for $\DeltaH$, can be traced back to the same origin, namely the saturation of the ALI. This is the only property a system is required to have in order for our result to apply. Very little is known about the explicit form of the modular Hamiltonians for the holographic examples mentioned above, so it is remarkable that they share this common property.
\\

Note that for the holographic examples described above, the ALI was assumed to be saturated for all considered $\sigma$ and $\lambda$. However, whether the ALI is saturated for a given value of $\lambda$ depends on the value of $\sigma$. If $\sigma$ is chosen too large the corresponding RT surfaces undergo a phase transition \cite{Headrick:2007km, Headrick:2010, Blanco:2013joa} that causes the ALI to be no longer saturated. Consequently, our result can only be applied to make statements for $\sigma$ sufficiently small
and $\lambda$ sufficiently close to the reference value\footnote{It depends on the chosen value of $\lambda$ for which $\sigma$ the phase transition of the RT surface occurs.} $\lambda_0$.
\\

We also need to stress that the saturation of the ALI inequality for the holographic situations discussed in \Sec \ref{sec: applications} is a large $N$
effect. Bulk quantum corrections to the RT formula are expected to lead to additional contributions to entanglement entropies in such a way that the ALI is no longer saturated \cite{Faulkner:2013ana}. So strictly speaking our result can only be used to show that $\DeltaH(A_\sigma,\lambda)$ or $\DeltaH(B_\sigma,\lambda)$ is in general non-linear in the respective $\tilde{\lambda}$ in the large $N$ limit.
By continuity, we expect this non-linearity to hold for finite $N$ as well.
\\

We emphasize once more that even though our result was mostly applied to
examples from AdS/CFT in this paper, it is not restricted to the holographic case. We only required the monotonicity \eqref{eq: monotonicity of Srel} of the RE and the first law of entanglement \eqref{eq: 1st law} -- which is a direct implication of the non-negativity of the RE -- to prove it. Both are known to be true for any quantum system. Therefore our result is an implication of well-established properties of the
RE and holds for generic quantum systems.
\\

The RE is a valuable object for studying modular Hamiltonians \cite{Blanco:2013lea, Faulkner:2016mzt, Ugajin:2016opf, Arias:2016nip, Blanco:2017akw} and offers prominent relations between modular Hamiltonians and entanglement entropies.
Our result is a further application of the RE that reveals such a relation. Unlike the first law of entanglement, which focuses on the first order contribution of $\tilde{\lambda}$ to $\DeltaH$, our result makes a statement about higher-order contributions in $\tilde{\lambda}$. The fact that the entanglement entropy plays a role for the higher-order contributions in $\tilde{\lambda}$ is a non-trivial observation that deserves further analysis. Possible future projects could be devoted to investigating whether
it is possible to find more concrete relations between entanglement entropy and higher-order $\tilde{\lambda}$ contributions to $\DeltaH$.
This will provide further progress towards understanding the properties of the
modular Hamiltonian in general QFTs.
\begin{acknowledgments}
We would like to thank Charles Melby-Thompson, Christian Northe and Ignacio Reyes for inspiring
discussions and Ren\'e Meyer and  Erik Tonni for fruitful conversations.
\end{acknowledgments}

\appendix
\section{Detailed Discussion of the Proof Presented in \Sec \ref{sec: gen EP}}
In \Sec \ref{sec: gen EP} we proved our main result for $S(A_\sigma,\lambda)\geq S(B_\sigma,\lambda)$, i.e.
\begin{equation}
	S(A_\sigma,\lambda)-S(B_\sigma,\lambda)
	=
	S(\Sigma,\lambda)\,.
\end{equation}
Here we show how the proof of this special case can be generalized to the situation
\begin{equation}
\label{eq: ALI appendix}
	|S(A_\sigma,\lambda)-S(B_\sigma,\lambda)|
	=
	S(\Sigma,\lambda)\,.
\end{equation}
We use the notation introduced in \Sec \ref{sec: gen EP}.
\\

First we note that the sign of $S(A_\sigma,\lambda)-S(B_\sigma,\lambda)$
does not change with $\sigma$, if \eqref{eq: ALI appendix} holds. For $S(\Sigma,\lambda)=0$ this is obvious. If the sign would change with $\sigma$ for $S(\Sigma,\lambda)\neq 0$, the continuity of $\sigma$ would imply that there is a $\sigma=\sigma'$ where $S(A_{\sigma'},\lambda)=S(B_{\sigma'},\lambda)$, which would lead to $S(\Sigma,\lambda)= 0$ and therefore contradict our assumption.
So we find that the sign of $S(A_\sigma,\lambda)-S(B_\sigma,\lambda)$ only changes in $\lambda$ and consequently
\begin{equation}
S(A_\sigma,\lambda)
	=
	S(B_\sigma,\lambda)
	\pm
	S(\Sigma,\lambda)\,,
\end{equation}
where only $\lambda$ dictates which sign in front of $S(\Sigma,\lambda)$
has to be chosen.

The next step is to distinguish the two situations $S(\Sigma,\lambda_0)\neq 0$
and $S(\Sigma,\lambda_0)=0$.
For $S(\Sigma,\lambda_0)\neq 0$ we can w.l.o.g.~assume $S(A_\sigma,\lambda_0)>	S(B_\sigma,\lambda_0)$. This inequality also holds for a small region around $\lambda_0$ which implies
\begin{equation}
	\Delta S'(A_\sigma,\lambda_0)
	=
	\Delta S'(B_\sigma,\lambda_0)
	+
	\Delta S'(\Sigma, \lambda_0)\,.
\end{equation}
By following the arguments of \Sec \ref{sec: gen EP} this leads to
\begin{equation}
\label{eq: appendix Srel of A}
	S_{rel}(A_{\sigma},\lambda)
	=
	\Delta S'(\Sigma,\lambda_0)\tilde{\lambda}
	\mp
	\Delta S(\Sigma,\lambda)
	+
	S_{rel}(B_{\sigma},\lambda)
\end{equation}
instead of \eqref{eq: SrelA i.t.o. SrelB for alpha} for $\sigma\in[\xi,\eta]$. Note that the sign in front of $\Delta S(\Sigma,\lambda)$ in \eqref{eq: appendix Srel of A} is the same for all $\sigma\in[\xi,\eta]$. Therefore the rest of the proof of our result
is analogous to the arguments presented in \Sec \ref{sec: gen EP}
below \eqref{eq: SrelA i.t.o. SrelB for alpha}.

For $S(\Sigma,\lambda_0)=0$ the non-negativity of the entanglement entropy
implies that $S(\Sigma,\lambda)$ takes its minimal value for $\lambda=\lambda_0$.
Therefore we find $S'(\Sigma,\lambda_0)=0$ and consequently
\begin{equation}
	\Delta S'(A_\sigma,\lambda_0)
	=
	\Delta S'(B_\sigma,\lambda_0)\,.
\end{equation}
This leads to
\begin{equation}
	S_{rel}(A_{\sigma},\lambda)
	=
	\mp
	\Delta S(\Sigma,\lambda)
	+
	S_{rel}(B_{\sigma},\lambda)
\end{equation}
instead of \eqref{eq: SrelA i.t.o. SrelB for alpha} for $\sigma\in[\xi,\eta]$. Just as for $S(\Sigma,\lambda)\neq 0$, the rest of the proof
can be formulated in an analogous way as in \Sec \ref{sec: gen EP}.
\bibliographystyle{JHEP}
\bibliography{mod_ham_lib}
\end{document}